\documentclass[prd,aps,nofootinbib,onecolumn]{revtex4}
\usepackage{epsfig,graphicx}
\usepackage{amscd}
\usepackage{amsmath}
\usepackage{graphicx}
\usepackage{amssymb}
\usepackage{float}
\usepackage{cancel}
\usepackage{axodraw4j}
\usepackage{color}

\newcommand{\be}{\begin{eqnarray}}
\newcommand{\ed}{\end{eqnarray}}

\begin{document}

\title{\bf Branching ratios and direct $CP$ asymmetries in $D\to PV$ decays}

\author{Qin Qin$^1$}\email{qqin@ihep.ac.cn}
\author{Hsiang-nan Li$^{2,3,4}$}\email{hnli@phys.sinica.edu.tw}
\author{Cai-Dian L\"u$^1$}\email{lucd@ihep.ac.cn}
\author{Fu-Sheng Yu$^{1,5,6}$}\email{yufsh@lzu.edu.cn}

\affiliation{$^{1}$Institute of High Energy Physics and Theoretical Physics
Center for Science  Facilities, Chinese Academy of Sciences, Beijing
100049, People's Republic of China,}

\affiliation{$^2$Institute of Physics, Academia Sinica, Taipei,
Taiwan 115, Republic of China,}

\affiliation{$^3$Department of Physics, National Tsing-Hua
University, Hsinchu, Taiwan 300, Republic of China,}

\affiliation{$^4$Department of Physics, National Cheng-Kung
University, Tainan, Taiwan 701, Republic of China,}

\affiliation{$^5$Laboratoire de l'Acc$\acute{e}$l$\acute{e}$rateur Lin$\acute{e}$aire, Universit$\acute{e}$ Paris-Sud 11, CNRS/IN2P3(UMR 8607) 91405 Orsay, France}

\affiliation{$^6$School of Nuclear Science and Technology, Lanzhou University, Lanzhou 730000, People's Republic of China}

\date{\today}
\begin{abstract}

We study the two-body hadronic $D\to PV$ decays, where $P$ ($V$)
denotes a pseudoscalar (vector) meson, in the factorization-assisted
topological-amplitude approach proposed in our previous work.
This approach is based on the factorization of
short-distance and long-distance dynamics into Wilson coefficients
and hadronic matrix elements of four-fermion operators, respectively,
with the latter being parametrized in terms of several
nonperturbative quantities. We further take into account the
$\rho$-$\omega$ mixing effect, which improves the global fit to the
branching ratios involving the $\rho^0$ and $\omega$ mesons.
Combining short-distance dynamics associated with
penguin operators and the hadronic parameters determined from the
global fit to branching ratios, we predict direct $CP$ asymmetries.
In particular, the direct $CP$ asymmetries in the
$D^0\to K^0\overline{K}^{*0},~\overline{K}^0K^{*0}$,
$D^+\to\pi^+\rho^0$, and $D_s^+\to K^+\omega,~K^+\phi$ decays are
found to be of ${\cal O}(10^{-3})$,
which can be observed at the LHCb or future Belle II experiment.
We also predict the $CP$ asymmetry observables of some neutral $D$ meson decays.

\end{abstract}


\maketitle

\section{Introduction}

Recent measurements of direct $CP$ asymmetries in two-body
hadronic $D$ meson decays have stimulated great theoretical
efforts on their study. The difference between the direct $CP$
asymmetries of the $D^0\to K^+ K^-$ and $D^0\to \pi^+\pi^-$ decays,
$\Delta A_{\rm CP}\equiv A_{\rm CP}(K^+K^-)-A_{\rm CP}(\pi^+\pi^-)=
[-0.82\pm0.21({\rm stat})\pm0.11({\rm syst})]\%$, was
observed by the LHCb \cite{arXiv:1112.0938} and confirmed
by other collaborations. For example, the CDF and Belle measurement gave
$\Delta A_{\rm CP}=[-0.62\pm0.21({\rm stat})\pm0.10({\rm syst})]\%$
\cite{CDF12} and $\Delta A_{\rm CP} = [-0.87 \pm 0.41 ({\rm stat}) \pm
0.06 ({\rm syst})]\%$ \cite{Belle12}, respectively.
The quantity $\Delta A_{\rm CP}$ is expected to be much
smaller in the Standard Model (SM), because the responsible penguin
contributions are suppressed by both the CKM matrix elements and the
Wilson coefficients \cite{Grossman:2006jg,Bigi:2011re},
$A_{\rm CP}\sim (|V_{cb}^*V_{ub}|/|V_{cs}^*V_{us}|)
(\alpha_s/\pi)\sim10^{-4}$.
The dramatic deviation of the data from the expectation has
been investigated in the SM and in new physics models
by employing different approaches.

To predict direct $CP$ asymmetries, a reliable evaluation of the penguin
contributions to two-body hadronic $D$ meson decays is necessary.
In Refs. \cite{Cheng:2012wr,Brod:2011re} the tree
amplitudes were determined by fitting the topology parametrization
to measured branching ratios, while the penguin amplitudes
were calculated in the QCD-improved factorization
\cite{BBNS99,BBNS01}. It has been noticed that the penguin
amplitudes derived from the QCD-improved factorization lead to a tiny
$\Delta A_{\rm CP}$ of order $10^{-5}$ \cite{Cheng:2012wr}. Allowing
the penguin amplitudes to be of the same order as the tree ones
discretionally, $\Delta A_{\rm CP}$ reaches $-0.13\%\sim O(10^{-3})$
\cite{Cheng:2012wr}. In another work \cite{Bhattacharya:2012ah} also
based on the topology parametrization, the penguin contribution via
an internal $b$ quark was identified as the major source of $CP$
violation, since it cannot be related to the tree amplitudes. This
penguin contribution, including its strong phase, was constrained by
the LHCb data and then adopted to predict direct $CP$ asymmetries of
other decay modes. Therefore, it is difficult to tell whether the
large $\Delta A_{\rm CP}\sim O(10^{-2})$ arise from new physics
\cite{Pirtskhalava:2011va,Feldmann:2012js,Chen:2012am,Altmannshofer:2012ur,
Giudice:2012qq,Hochberg:2011ru,Isidori:2011qw,Rozanov:2011gj},
if one follows the approaches in the literature.

To estimate the penguin contribution precisely, we have proposed
a theoretical framework for two-body hadronic $D$ meson decays,
named as the factorization-assisted topological-amplitude (FAT) approach
\cite{Li:2012cfa}, which combines the conventional naive factorization
hypothesis and topological-amplitude parametrization. It is based on
the factorization of short-distance (long-distance) dynamics into Wilson
coefficients (hadronic matrix elements of four-fermion operators, i.e.,
topological amplitudes). Because
of the small charm quark mass just above 1 GeV, a perturbation theory for
the hadronic matrix elements may not be reliable. The idea is to identify
as complete as possible the important sources of nonperturbative dynamics
in the hadronic matrix elements and parametrize them in the framework of the
factorization hypothesis. Fitting our parametrization
to abundant data of $D$ meson decay branching ratios, all the
nonperturbative parameters can be determined. Once the nonperturbative
parameters have been determined, the replacement of the Wilson
coefficients works for estimating the penguin contributions. For
those penguin amplitudes, which cannot be related to tree amplitudes
through the above replacement, we have shown that they are either
factorizable or suppressed by the helicity conservation. If they are
factorizable, such as the scalar penguin annihilation contribution, data
from other processes can be used for their determination. We are
then able to predict the direct $CP$ asymmetries in $D$ meson decays
without ambiguity.

The FAT approach has been applied to the study of the $D\to PP$ decays
\cite{Li:2012cfa}, where $P$ represents a pseudoscalar meson. It has
been shown that our framework greatly improves the global fit to the
measured $D\to PP$ branching ratios. In particular, we have
obtained $\Delta A_{\rm CP}=-1.00\times 10^{-3}$, which discriminates
the opposite postulations on large (small) direct $CP$ asymmetries in
singly Cabibbo-suppressed $D$ meson decays \cite{GG89}
(\cite{pietro}). After the publication of our work, the LHCb collaboration
updated the data\cite{LHCbBB},
\begin{eqnarray}
\Delta A_{\rm CP} = [-0.34 \pm 0.15({\rm stat}) \pm 0.10({\rm syst})]\%,\label{update}
\end{eqnarray}
where the central value is lower than the previous one. Two sources of
$D$ meson production have been employed by the LHCb collaboration in the
measurements of $\Delta A_{\rm CP}$: the $D^{*+} \to D^0\pi^+$ channel with the
flavor of the neutral $D$ meson being determined by the emitted pion and
semileptonic $b$-hadron decays where the flavor of the neutral $D$ meson
is tagged by the accompanying charged lepton. The former,
from more data collected in the fall of 2011, led to
Eq.~(\ref{update}) with lower statistical uncertainty.
The latter from almost one-third of the samples of the $D^*$ analysis gave
\cite{LHCbB}
\begin{eqnarray}
\Delta A_{\rm CP} = [+0.49 \pm 0.30({\rm stat}) \pm 0.14({\rm syst})]\%.
\end{eqnarray}
The sign flip of the central value indicates that the direct $CP$ asymmetry
in the $D^0\to K^+ K^-$, $\pi^+\pi^-$ decays may be
small, so it could fluctuate into negative or positive values.

In this paper we shall extend the FAT approach to the $D\to PV$ decays with $V$
denoting a vector meson. Their data of branching ratios are also abundant
enough for fixing nonperturbative parameters, and their direct $CP$
asymmetries are of great phenomenological importance and interest.
Compared to Ref. \cite{Li:2012cfa}, we further take into account the
$\rho$-$\omega$ mixing effect, which improves the global fit to the
branching ratios involving the $\rho^0$ and $\omega$ mesons. It will be
shown that the measured branching ratios of the $D_s^+$ decays into
$K^+\overline K^{*0}$, $\overline K^0 K^{*+}$, $\pi^+\rho^0$, and $\pi^+\omega$,
which could not be accommodated simultaneously in the diagrammatic
approach \cite{Cheng:2010ry}, are explained. This overall improvement
between the predictions and the data is attributed
to the SU(3) symmetry breaking effects included in our
topological-amplitude parametrization. Besides, the direct $CP$
asymmetries in the $D^0\to K^0\overline{K}^{*0},~\overline{K}^0K^{*0}$,
$D^+\to\pi^+\rho^0$, and $D_s^+\to K^+\omega,~K^+\phi$
modes reach $10^{-3}$, which can be observed
at LHCb or future Belle II. We also calculate
the $CP$ asymmetry observables of some neutral $D$ meson decays.
Our predictions presented in this work would help analyze
$CP$ asymmetries in three-body $D$ meson decays. For example,
the result for the $D^0\to\pi^0\rho^0$ mode is relevant
to the $D^0 \to \pi^+\pi^-\pi^0$ channel.

In Sec.~II we construct our parametrization of the tree contributions
to the $D\to PV$ branching ratios in the FAT approach. In Sec.~III the
penguin contributions from the operators $O_{3-6}$, from $O_{1,2}$
through the quark loops, and from the magnetic penguin $O_{8g}$ are
formulated. The direct $CP$ asymmetries in the $D\to PV$ decays are
then predicted. Section~IV is the conclusion. We discuss the scalar
penguin contributions in Appendix~A and the $\rho$-$\omega$ mixing
in Appendix~B.

\section{Branching Ratios}

In the FAT approach the hadronic matrix elements of the four-fermion
operators, including the emission, $W$-annihilation, and $W$-exchange
amplitudes, are parametrized into the magnitudes $\chi$'s and
the strong phases $\phi$'s. An important ingredient is the
Glauber strong phase factor \cite{CL11} associated with a pion
in the nonfactorizable annihilation amplitudes, which might originate
from the unique role of the pion as a Nambu-Goldstone boson and a
quark-antiquark bound state simultaneously. The Glauber phase
modifies the relative angle and the interference between the
annihilation and emission amplitudes involving pions. The predicted
$D^0\to\pi^+\pi^-$ ($D^0\to K^+K^-$) branching ratio is then reduced
(enhanced), and the the long-standing puzzle related to these
branching ratios \cite{BR10,Cheng:2010ry} is resolved. In this work,
we only consider the tree contributions to the branching ratios and
neglect the penguin ones which are suppressed by Wilson coefficients
and CKM matrix elements.

\subsection{Parametrization of tree amplitudes}

\begin{figure}[!ht]
  \begin{center}
  \includegraphics[scale=0.5]{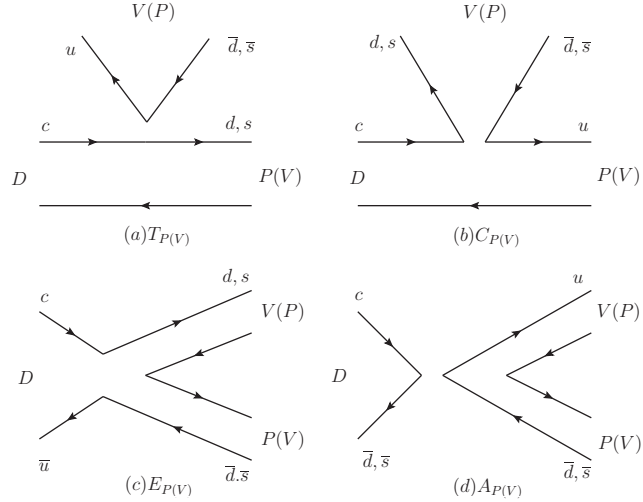}
  \vspace{-7cm}
  \caption{Eight topological diagrams contributing to the
  $D\to PV$ decays with (a) the color-favored tree amplitude
  $T_{P(V)}$, (b) the color-suppressed tree amplitude $C_{P(V)}$,
  (c) the $W$-exchange amplitude $E_{P(V)}$, and (d) the $W$-annihilation
  amplitude $A_{P(V)}$.}\label{fig:I}
  \end{center}
\end{figure}

In this subsection we parametrize the tree
contributions which dominate the $D\to PV$ branching ratios.
The relevant effective weak Hamiltonian is given by
\begin{equation}
\mathcal{H}_{eff}=\frac{G_F}{\sqrt{2}}V_{CKM}[C_1(\mu)O_1(\mu)+C_2(\mu)O_2(\mu)],
\end{equation}
where $G_F$ is the Fermi coupling constant, $V_{CKM}$ represents the product
of the corresponding CKM matrix elements, and $C_{1,2}$ are the
Wilson coefficients. The current-current operators are defined by
\begin{equation}\begin{split}
&O_1=(\overline{u}_{\alpha}q_{2\beta})_{V-A}(\overline{q}_{1\beta}c_{\alpha})_{V-A},\\
&O_2=(\overline{u}_{\alpha}q_{2\alpha})_{V-A}(\overline{q}_{1\beta}c_{\beta})_{V-A},
\end{split}\end{equation}
with $q_{1,2}$ being the $d$ or $s$ quark, $\alpha,\beta$ being the color
indices, and $(\overline{q}q')_{V-A}$ representing $\overline{q}\gamma_\mu(1-\gamma_5)q'$.
The relevant eight topological diagrams are displayed in Fig.~\ref{fig:I},
where $T_{P(V)}$ represents the color-favored tree amplitude with the
$D\to P(V)$ transition, $C_{P(V)}$ represents the
color-suppressed tree amplitude with the $D\to P(V)$
transition, $E_{P(V)}$ represents the $W$-exchange amplitude with
the pseudoscalar (vector) meson containing the antiquark from the
weak vertex, and $A_{P(V)}$ represents the $W$-annihilation amplitude
with the pseudoscalar (vector) meson containing the antiquark from
the weak vertex.

For the emission type, we ignore the nonfactorizable contributions to
the color-favored amplitudes because the factorizable ones dominate.
The amplitudes $T_P$ and $C_P$ are formulated as \cite{Li:2012cfa}
\be
T_P(C_P)={G_F\over\sqrt{2}}V_{CKM} a_1(\mu)\left( a_2^P(\mu)\right) f_V m_V
F_1^{DP}(m_V^2)2(\varepsilon_V\cdot p_D),\label{fdp}
\ed
where $f_V$ ($m_V$, $\varepsilon_V$) is the decay constant (mass,
polarization vector) of the vector meson,
$F_1^{DP}$ is the $D\to P$ transition form
factor, and $p_D$ is the $D$ meson momentum. The amplitudes $T_V$
and $C_V$ are formulated as
\be
T_V(C_V)={G_F\over\sqrt{2}}V_{CKM} a_1(\mu)\left(a_2^V(\mu)\right)f_P m_V
A_0^{DV}(m_P^2)2(\varepsilon_V\cdot p_D),
\ed
where $f_P$ is the decay constant of the pseudoscalar meson and
$A_0^{DV}$ is the $D\to V$ transition form factor. The associated
scale-dependent Wilson coefficients $a_1$ and $a_2^{P,V}$ are given by
\be
a_1(\mu)&=&C_2(\mu)+{C_1(\mu)\over N_C},\nonumber\\
a_2^{P(V)}(\mu)&=&C_1(\mu)+C_2(\mu)
\left({1\over N_C}+\chi_{P(V)}^Ce^{i\phi_{P(V)}^C}\right),\label{a12}
\ed
with $N_C$ being the number of colors. The parameters $\chi_{P,V}^C$ and
$\phi_{P,V}^C$ describe the magnitudes and the strong phases of the nonfactorizable
contributions in the color-suppressed amplitudes, since final-state interaction (FSI) and resonance
effects cannot be neglected in $D$ meson decays. We set the scale of the Wilson
coefficients to the energy release in individual decay modes as suggested by the
perturbative QCD (PQCD) approach \cite{Li:2001pqcd}:
it depends on masses of final states and on the scale $\Lambda$ that
characterizes the soft degrees of freedom in the $D$ meson \cite{Li:2012cfa},
\be
\mu=\sqrt{\Lambda m_D(1-r_{V(P)}^2)},
\ed
$r_{V(P)}=m_{V(P)}/m_D$ being the mass ratio of the vector (pseudoscalar)
meson emitted from the weak vertex over the $D$ meson. The evolution of
the Wilson coefficients for $c$ quark decays can be found in Ref.
\cite{Li:2012cfa}.

Because the factorizable contributions to the annihilation-type
amplitudes are down by helicity suppression\cite{Hoang:1995ns}, only the
nonfactorizable contributions are considered.
The $W$-exchange and $W$-annihilation amplitudes are parametrized as
\be\label{eq:E}
E_{P,V}&=&{G_F\over\sqrt{2}}V_{CKM}C_2(\mu)\chi^E_{q(s)}e^{i\phi^E_{q(s)}}
f_Dm_D {f_P\over f_\pi}{f_V\over f_\rho}(\varepsilon_V\cdot p_D),\\
A_{P,V}&=&{G_F\over\sqrt{2}}V_{CKM}C_1(\mu)\chi^A_{q(s)}e^{i\phi^A_{q(s)}}
f_Dm_D {f_P\over f_\pi}{f_V\over f_\rho}(\varepsilon_V\cdot p_D),\label{eq:A}
\ed
where $f_D$, $f_{\pi}$, and $f_{\rho}$ are the decay
constants of the $D$ meson, $\pi$ meson, and $\rho$ meson, respectively.
The parameters $\chi^{E,A}_{q,s}$ and $\phi^{E,A}_{q,s}$
characterize the strengths and the strong phases of the corresponding amplitudes,
with the subscripts $q$ and $s$ differentiating the strongly produced light-quark
($u$ or $d$) and strange-quark pair. The
ratios over $f_{\pi}$ and $f_\rho$ in Eqs.~(\ref{eq:E}) and (\ref{eq:A}) take into
account the $SU(3)$ breaking effects from the decay constants. As in the
emission-type amplitudes, the scale of the Wilson coefficients,
\be
\mu=\sqrt{\Lambda m_D(1-r_P^2)(1-r_V^2)},
\ed
also depends on the initial- and final-state masses.

As shown above, we have followed the parametrization for the
$D\to PP$ decays \cite{Li:2012cfa} by considering the nonfactorizable
amplitudes $\chi_q$ and $\chi_s$ in this work. Note that $\chi_P$ and $\chi_V$
were adopted in Ref. \cite{Cheng:2010ry}, which describe the nonfactorizable
contributions with the spectator antiquark going into the $P$ and $V$
mesons, respectively. However, as $\chi_P$ and $\chi_V$ appear together
in some $D\to PV$ modes, such as $D^+\to \pi^+\omega$, their difference
reflects the isospin symmetry breaking, which ought to be tiny. Certainly,
they do not always appear together. For example, only $\chi_P$ appears
in the $D^0\to \pi^+\rho^-$ decay. Viewing that $\chi_P$ and $\chi_V$
may violate the isospin symmetry, we prefer $\chi_q$ and $\chi_s$, for
which the difference reflects the SU(3) symmetry breaking that could be
significant. It turns out that the parametrization with $\chi_q$ and
$\chi_s$ has a lower $\chi^2$ in the global fit than the parametrization
with $\chi_P$ and $\chi_V$ does. That is, the SU(3) symmetry breaking is
more crucial than the isospin symmetry breaking in $D$-meson decays.

It was proposed in Ref. \cite{Li:2011gf} that a kind of soft gluons, named
the Glauber gluons, exist in two-body heavy meson decays, which may
lead to additional strong phases in the nonfactorizable amplitudes.
The multiple Fock states of a pion have been proposed to reconcile its
simultaneous roles as a $q\bar q$ bound state and a Nambu-Goldstone boson \cite{NS08}.
It was then speculated that the Glauber effect becomes significant due
to the huge soft cloud formed by higher Fock states of a pion \cite{Li:2011gf}.
According to Ref. \cite{Li:2012cfa}, we multiply a phase factor $\exp(iS_\pi)$ to the
nonfactorizable annihilation-type amplitudes, as a pion is involved in the
final state, while leaving the emission-type amplitudes unchanged, in which
the factorizable contributions usually dominate.
In summary, our parametrization of the $D\to PV$
decays is composed of 14 global free parameters: the
soft scale $\Lambda$; the magnitudes of the nonfactorizable amplitudes,
$\chi^C_{P,V}$ and $\chi^{E,A}_{q,s}$; the strong phases of the
nonfactorizable amplitudes, $\phi^C_{P,V}$ and $\phi^{E,A}_{q,s}$;
and the Glauber phase $S_\pi$. Compared to the $D\to PP$ analysis
\cite{Li:2012cfa}, there are only two more free parameters.

\subsection{Numerical analysis}

The partial decay width of a $D\to PV$ mode
is expressed as
\begin{equation}
\Gamma(D\to PV)={|\vec p_V|\over8\pi m_D^2}\sum_{pol}|\mathcal{A}|^2,\label{d1}
\end{equation}
or equivalently as
\begin{equation}
\Gamma(D\to PV)={|\vec p_V|^3\over8\pi m_V^2}|\tilde{\mathcal{A}}|^2,\label{d2}
\end{equation}
which are related to each other via $\mathcal{A}=\tilde{\mathcal{A}}(\varepsilon\cdot p_D)$
and the summation over the polarization states of the vector boson,
$\sum_{pol}|\varepsilon\cdot p_D|^2=(m_D^2/ m_V^2)|\vec p_V|^2$. Note
that only the longitudinal polarization state of the vector meson
contributes to the $D\to PV$ decays. We perform the global fits based
on the above two formulas and find the same solutions. This is in
contrast to the observation in Ref. \cite{Cheng:2010ry}, where different
solutions were obtained from Eqs.~(\ref{d1}) and (\ref{d2}). For the
decay constants of the pseudoscalar and vector mesons and the $D\to P$
transition form factors $F_1^{DP}$ in Eq.~(\ref{fdp}), we take the same
values as in Ref. \cite{Fusheng:2011tw}. The $D\to V$ transition form
factors $A_0^{DV}(q^2)$'s have been calculated with poor precision:
their value at $q^2=0$ range from about 0.6 to 0.8 \cite{formfactor}
and are chosen as in Table~\ref{formfactorset}. Our fits include all
the channels with measured branching ratios except $D_s^+\to\eta'\rho^+$,
i.e., 33 experimental data of branching ratios in total. The
$D_s^+\to\eta'\rho^+$ mode is excluded for the following reason. It is
the only $\eta'$-involved decay with a measured branching ratio, and
the input of the $D_s^+\to\eta'$ transition form factor is uncertain,
for which the variation easily changes the fit to this mode. Therefore,
its data, with less satisfactory quality, does not constrain the relevant
parameters effectively.

\begin{table}[!htbh]
  \centering
  \caption{Values of $D\to V$ transition form factors $A_0^{DV}(0)$.}\label{formfactorset}
  \begin{tabular}[t]{ccccc}\hline\hline
  $D\to\rho$ & $D\to K^*$ & $D\to\omega$ & $D_s\to K^*$ & $D_s\to\phi$\\
  0.76 & 0.73 & 0.70 & 0.76 & 0.78\\
  \hline\hline
  \end{tabular}
\end{table}

The global fit leads to the nonperturbative parameters
\begin{equation}\label{eq:para}\begin{split}\
&\Lambda=0.44\;\;{\rm GeV},~S_{\pi}=-0.96,\\
&\chi_P^C=-0.40,~\phi_P^C=-0.53,~\chi_V^C=-0.53,~\phi_V^C=-0.25,\\
&\chi^E_q=0.25,~\phi^E_q=1.73,~\chi^A_q=0.11,~\phi^A_q=-0.35,\\
&\chi^E_s=0.29,~\phi^E_s=3.11,~\chi^A_s=0.10,~\phi^A_s=1.60,
\end{split}\end{equation}
with the fitted $\chi^2=2.8$ per degree of freedom. Since the weak phases
associated with the tree contributions are tiny, roughly the same
branching ratios will be obtained, if the strong phases in
Eq.~(\ref{eq:para}) flip the sign. We select the above outcomes to keep the
strong phases of the emission-type amplitudes in consistence with those
in Ref. \cite{Li:2012cfa}. The value of $\Lambda$ is in the correct order of
magnitude for characterizing the soft degrees of freedom in the $D$
meson and close to that derived from the $D\to PP$ fit
\cite{Li:2012cfa}. The Glauber phase $S_\pi$ is not very different
from what was obtained in the $D\to PP$ analysis \cite{Li:2012cfa} and is
consistent with the value extracted from the data for the direct $CP$ asymmetries
in the $B\to\pi K$ decays \cite{Li:2011gf}.

The branching ratios of the Cabibbo-favored, singly Cabibbo-suppressed,
and doubly Cabibbo-suppressed $D\to PV$ decays corresponding to the
parameters in Eq.~(\ref{eq:para}) are listed in
Tables~\ref{1tb:AbrA}, \ref{2tb:BbrB}, and \ref{3tb:CbrC},
respectively. Our results are also compared with the experimental data
\cite{2012PDG} and with those from other theoretical approaches, such as the
fit based on the diagrammatic approach \cite{Cheng:2010ry}, the
calculations including the FSI effects of nearby resonances
\cite{Buccella:1995fsi}, and the combination of the generalized
factorization and the pole model \cite{Fusheng:2011tw}. Our results in
the column Br(FAT) basically agree with the data. Note that the branching
ratios $Br(D_s^+\to \eta'\rho^+)=(12.5\pm2.2)\%$ listed in the PDG2012 \cite{2012PDG}
was from an old measurement \cite{CLEO:1998eta'}. It was then questioned for
exceeding the inclusive $\eta'$ fraction $(11.7\pm1.8)\%$ \cite{2012PDG},
which includes all $\eta'$ involved modes. This controversy was resolved
by the recent CLEOc measurement with the branching fraction
$Br(D_s^+\to \eta'\rho^+)=(5.6\pm1.1)\%$ \cite{Onyisi:2013bjt}, which is
closer to our prediction.

\begin{table}[!htbh]
  \centering
  \caption{Branching ratios for the Cabibbo-favored $D\to PV$ decays
  in units of percentage. Our results without (FAT) and with the $\rho$-$\omega$ mixing
  (FAT[mix]) are compared to the experimental data \cite{2012PDG}, the fitted results
  from the diagrammatic approach \cite{Cheng:2010ry}, the results including the
  FSI effects \cite{Buccella:1995fsi}, and the calculations from the combination of
  the generalized factorization and the pole model \cite{Fusheng:2011tw}.
  The involved amplitudes of the decays are also shown,
  with those outside the parentheses being dominant.}\label{1tb:AbrA}
  \begin{tabular}[t]{cccccccc}\hline\hline
  Modes & Amplitudes & Br(FSI) & Br(diagrammatic)& Br(pole) & Br(FAT)&Br(FAT[mix]) & Br(exp)\\\hline
  $D^0\to \pi^+K^{*-}$ & $T_V,(E_P)$ &4.69&$5.91\pm0.70$&$3.1\pm1.0$& 6.21 &6.09& $5.44^{+0.70}_{-0.53}$ \\
  $D^0\to \pi^0\overline K^{*0}$ & $C_P$,($E_P$) &3.49&$2.82\pm0.34$&$2.9\pm1.0$& 3.42 &3.25& $3.44\pm0.35$ \\
  $D^0\to \overline{K}^0\rho^0$ & $C_V$,($E_V$) &0.88&$1.54\pm1.15$&$1.7\pm0.7$& 1.31 &1.17& $1.26^{+0.14}_{-0.16}$ \\
  $D^0\to \overline{K}^0\omega$ & $C_V$,($E_V$) &2.16&$2.26\pm1.38$&$2.5\pm0.7$& 2.26 &2.22 &$2.22\pm0.12$ \\
  $D^0\to \overline{K}^0\phi$ & $E_P$ &0.90&$0.868\pm0.139$&$0.8\pm0.2$& 0.800 &0.800& $0.834\pm0.074$ \\
  $D^0\to K^-\rho^+$ & $T_P$,($E_V$) &11.19&$10.8\pm2.2$&$8.8\pm2.2$& 9.6 &9.6& $10.8\pm0.7$ \\
  $D^0\to \eta\overline K^{*0}$ & $C_P$,($E_P$,$E_V$) &0.51&$0.96\pm0.32$&$0.7\pm0.2$& 0.55 &0.57& $0.96\pm0.30$ \\
  $D^0\to \eta'\overline K^{*0}$ & $C_P$,($E_P$,$E_V$) &0.005&$0.012\pm0.003$&$0.016\pm0.005$& 0.018 &0.018&$<0.11$ \\
  $D^+\to \pi^+\overline K^{*0}$ & $T_V$,$C_P$ &0.64&$1.83\pm0.49$&$1.4\pm1.3$& 1.70 &1.70& $1.51\pm0.16$ \\
  $D^+\to \overline{K}^0\rho^+$ & $T_P$,$C_V$ &11.77&$9.2\pm6.7$&$15.1\pm3.8$& 6.4 &6.0& $9.6\pm2.0$ \\
  $D_s^+\to \pi^+\rho^0$ & $A_P$,$A_V$ &0.080&&$0.4\pm0.4$& 0 &0.004& $0.020\pm0.012$ \\
  $D_s^+\to \pi^+\omega$ & $A_P$,$A_V$ &0.0&&0& 0.30 &0.26& $0.25\pm0.07$ \\
  $D_s^+\to \pi^+\phi$ & $T_V$ &2.89&$4.38\pm0.35$&$4.3\pm0.6$& 3.4 & 3.4 & $4.5\pm0.4$ \\
  $D_s^+\to \pi^0\rho^+$ & $A_P$,$A_V$ &0.080&&$0.4\pm0.4$& 0 &0&  \\
  $D_s^+\to K^+\overline K^{*0}$ & $C_P$,($A_V$) &3.86&&$4.2\pm1.7$& 4.08 &4.07& $3.95\pm0.2$ \\
  $D_s^+\to \overline{K}^0K^{*+}$ & $C_V$,($A_P$) &3.37&&$1.0\pm0.6$& 2.5 &3.1& $5.4\pm1.2$ \\
  $D_s^+\to \eta\rho^+$ & $T_P$,($A_P$,$A_V$) &9.49&&$8.3\pm1.3$& 8.2 &8.8& $8.9\pm0.8$ \\
  $D_s^+\to \eta'\rho^+$ & $T_P$,($A_P$,$A_V$) &2.61&&$3.0\pm0.5$& 1.7 & 1.6 & $5.6\pm1.1$\footnote{data from Ref. \cite{Onyisi:2013bjt}} \\
  \hline\hline
  \end{tabular}
\end{table}

It was noticed in Refs. \cite{Fusheng:2011tw,Buccella:1995fsi} that
the prediction for the $D_s^+\rightarrow \pi^+\rho^0$ branching ratio
is much larger than the data, while the $D_s^+\rightarrow \pi^+\omega$
branching ratio, predicted to be zero,
is sizable in experiments. The inconsistence observed in Refs.
\cite{Fusheng:2011tw,Buccella:1995fsi} was explained via the topological
amplitudes of these two modes:
\be
\mathcal{A}(D_s^+\to \pi^+\rho^0)&=&{1\over\sqrt{2}}(A_P-A_V),\\
\mathcal{A}(D_s^+\to \pi^+\omega)&=&{1\over\sqrt{2}}(A_P+A_V).
\ed
The factorizable $W$-annihilation contributions $A_P^f$ and $A_V^f$ obey
$A_P^f=-A_V^f$, which holds in the pole-dominant model \cite{Fusheng:2011tw}
because of the antisymmetric space wave function of the two $P$-wave final
states and can also be derived in the PQCD approach \cite{Lu:2000hj}.
Then the two contributions are constructive in the $\pi^+\rho^0$ mode and
destructive in the $\pi^+\omega$ mode, contrary to the implication of the
data. In our approach only the nonfactorizable contributions are
considered due to the helicity suppression of the factorizable ones
as shown in Eq.~(\ref{eq:A}), such that the relation $A_P= A_V$ leads to
the vanishing $D_s^+\rightarrow \pi^+\rho^0$ branching ratio [see the value in
the column Br(FAT) of Table~\ref{1tb:AbrA}]. The difference between our
prediction and those in Refs. \cite{Fusheng:2011tw,Buccella:1995fsi} for
the $D_s^+\to \pi^0\rho^+$ branching ratio can be understood in the same way.

In the diagrammatic approach \cite{Cheng:2010ry}, where the global
fit was performed only for the Cabibbo-favored modes with the flavor SU(3)
symmetry, it is impossible to find a reasonable solution to the
$D_s^+\to \pi^+\rho^0$, $\pi^+\omega$, $K^+\overline K^{*0}$, and
$\overline K^0 K^{*+}$ data simultaneously. Besides, the fit in Ref.
\cite{Fusheng:2011tw} indicated that the $D_s^+\to \overline K^0 K^{*+}$
branching ratio is much lower than the $D_s^+\to K^+\overline K^{*0}$ one.
This is also the case observed in the naive factorization,
because of the form factor relation $F^{DK}_1\approx A^{DK^*}_0$,
but with the decay constant $f_K<f_{K^*}$. Note that
the $D_s^+\to \pi^+\rho^0$ and $\pi^+\omega$ decays involve the
$W$-annihilation amplitudes with strongly produced light-quark pairs,
while the $D_s^+\to K^+\overline K^{*0}$ and $\overline K^0 K^{*+}$ decays
involve strongly produced strange-quark pairs. Since we have included
the significant SU(3) breaking effects from the nonfactorizable
contributions in Eq.~(\ref{eq:A}), better agreement between the predictions
and the data for the above modes has been attained in our global fit.

It is seen that some $\omega$-involved branching ratios have been overestimated
compared to the corresponding $\rho^0$-involved ones. For example, the
predicted $D^0\to\pi^0\omega$ branching ratio in the column Br(FAT) of
Table~\ref{2tb:BbrB} exceeds the upper bound from the experimental value,
$Br(D^0\to \pi^0\omega)<0.26\times10^{-3}$, while the $D^0\to\pi^0\rho^0$
one is slightly lower. The predicted $D^+\to\pi^+\rho^0$ and
$D_s^+ \to K^+\rho^0$ branching ratios are also lower than the data, while
the $D^+\to\pi^+\omega$ and $D_s^+ \to K^+\omega$ ones may be overestimated.
The above observation implies that the inclusion of
the $\rho$-$\omega$ mixing effect may improve the consistency between the predictions
and the data for these decays. We notice the similar pattern in the $D\to PP$ analysis
\cite{Li:2012cfa}: for the $D\to\pi^0 K$ decays, for which the branching ratios
are larger, our predictions are consistent with the data. For the $D^0\to\pi^0\pi^0$ decay,
for which the branching ratio is smaller, it was underestimated. Taking into account the
$\pi$-$\eta$-$\eta'$ mixing, for which the mixing matrix element $M_{12}$ is negative
\cite{Qian:2009dc}, the predicted $D^0\to\pi^0\pi^0$ branching ratio
can be increased and match the data better. We point out that $M_{12}$ is negative
in the $\rho$-$\omega$ mixing \cite{Qian:2009dc}, so its effect on the $\omega$-involved
modes is opposite and in the desired tendency. It is intriguing that both the $D\to PP$
and $D\to PV$ decays exhibit the meson mixing mechanism.

Motivated by the above argument, we include the
$\rho$-$\omega$ mixing defined by
\begin{equation}\begin{split}\label{eq:mixing}
&|\rho^0\rangle=|\rho^0_I\rangle-\epsilon|\omega_I\rangle,\\
&|\omega\rangle=\epsilon|\rho^0_I\rangle+|\omega_I\rangle,\\
\end{split}\end{equation}
up to ${\cal O}(\epsilon^2)$ corrections, where
$|\rho^0_I\rangle$ and $|\omega_I\rangle$ denote the isospin eigenstates.
The decay constants of the $|\rho^0_I\rangle$ and $|\omega_I\rangle$ states
are related to those of the physical states in Appendix B, through
the evaluation of the $V^0\to e^+e^-$ decay width.
Choosing the mixing angle $\epsilon=0.12$, which
is reasonable viewing the large uncertainty of this parameter \cite{OTW97}, we obtain another
set of nonperturbative parameters,
\begin{equation}\label{eq:paras}\begin{split}\
&\Lambda=0.44\;\;{\rm GeV},~S_{\pi}=-0.85,\\
&\chi_P^C=-0.40,~\phi_P^C=-0.53,~\chi_V^C=-0.63,~\phi_V^C=-0.42,\\
&\chi^E_q=0.26,~\phi^E_q=1.74,~\chi^A_q=0.17,~\phi^A_q=-0.77,\\
&\chi^E_s=0.29,~\phi^E_s=3.10,~\chi^A_s=0.10,~\phi^A_s=1.61,
\end{split}\end{equation}
with the fitted $\chi^2=2.3$ per degree of freedom. The corresponding
$D\to PV$ branching ratios are listed in the column Br(FAT[mix]) of
Tables~\ref{1tb:AbrA}, \ref{2tb:BbrB}, and \ref{3tb:CbrC}. After
including the mixing, the predicted $D^0\to \pi^0\omega$ branching
ratio is reduced to $0.18\times10^{-3}$ mainly due to the lower
$\omega$ meson decay constant and is below the observed upper bound.
The branching ratios of most other $\omega$-involved modes are also
decreased considerably. On the contrary, the branching ratios of most
$\rho^0$-involved modes are enhanced, since the $\rho$ meson decay
constant is increased. However, some of them are lowered, such as the
branching ratios of $D^0\to \overline{K}^0\rho^0$ and $D^0\to \eta\rho^0$ .
The mixing effect has only a minor correction to the $\rho^0$ meson
decay constant, which is overcome by the changes of the parameters in
Eq.~(\ref{eq:paras}). Note that the $D^+\to\pi^+\omega$ branching ratio
around $8.0\times 10^{-4}$ is still higher than the experimental upper
bound $3.4\times 10^{-4}$ even after including the $\rho$-$\omega$
mixing. With the very limited number of free parameters in our
global fit, this outcome is acceptable.

To improve the global fit, we can include the nonfactorizable
contributions to the amplitude $T$ or the factorizable contributions
to the amplitudes $E$ and $A$, both of which are expected to be small
and have been neglected. With four more free parameters introduced in
each case, the $\chi^2$ is reduced from 44.4 to 36.6 and 36.3,
respectively. The additional contributions turn out to be tiny and
change the results for the branching ratios by only about $7\%$. The
original parameters remain almost the same, implying that the
additionally introduced parameters are indeed less important.
Improvement can also be achieved by tuning the inputs of the form
factors and the mixing angle, but it will not be pursued in this paper.

\begin{table}[!htbh]
  \centering
  \caption{Same as Table~\ref{1tb:AbrA} for the singly Cabibbo-suppressed
  $D\to PV$ decays in units of $10^{-3}$.}\label{2tb:BbrB}
  \begin{tabular}[t]{cccccccc}\hline\hline
  Modes & Amplitudes & Br(FSI) & Br(diagrammatic)&Br(pole) & Br(FAT) &Br(FAT[mix])& Br(exp)\\\hline
  $D^0\to \pi^+\rho^-$ & $T_V,(E_P)$ &6.5&$3.92\pm0.46$&$3.5\pm0.6$& 4.74 &4.66& $4.96\pm0.24$ \\
  $D^0\to \pi^0\rho^0$ &  $C_P$,$C_V$,($E_P$,$E_V$) &1.7&$2.96\pm0.98$&$1.4\pm0.6$& 3.55 &3.83& $3.72\pm0.22$ \\
  $D^0\to \pi^0\omega$ & $C_P$,$C_V$,($E_P$,$E_V$) &0.08&$0.10\pm0.18$&$0.08\pm0.02$& 0.85 &0.18 & $<0.26$ \\
  $D^0\to \pi^0\phi$ & $C_P$ &1.1&$1.22\pm0.08$&$1.0\pm0.3$& 1.11 &1.11& $1.31\pm0.10$ \\
  $D^0\to \pi^-\rho^+$ & $T_P$,($E_V$) &8.2&$8.34\pm1.69$&$10.2\pm1.5$& 10.2 &10.0& $9.8\pm0.4$ \\
  $D^0\to K^+K^{*-}$ & $T_V$,($E_P$) &2.8&$1.99\pm0.24$&$1.6\pm0.3$& 1.72& 1.73& $1.56\pm0.12$ \\
  $D^0\to K^0\overline{K}^{*0}$ & $E_P$,$E_V$ &0.99&$0.29\pm0.22$&$0.16\pm0.05$& 1.1 & 1.1 & $<1$  \\
  $D^0\to \overline{K}^0K^{*0}$ & $E_P$,$E_V$ &0.99&$0.29\pm0.22$&$0.16\pm0.05$& 1.1 & 1.1 & $<0.56$  \\
  $D^0\to K^-K^{*+}$ & $T_P$,($E_V$) &4.5&$4.25\pm0.86$&$4.7\pm0.8$ & 4.37&4.37& $4.38\pm0.21$ \\
  $D^0\to \eta\rho^0$ & $C_P$,$C_V$,($E_P$,$E_V$) &0.24&$1.11\pm0.86$&$0.05\pm0.01$& 0.54 &0.45&  \\
  $D^0\to \eta\omega$ & $C_P$,$C_V$,($E_P$,$E_V$) &1.9&$3.08\pm1.42$&$1.2\pm0.3$& 2.4 &2.0&  \\
  $D^0\to \eta\phi$ & $C_P$,($E_P$,$E_V$) &0.57&$0.31\pm0.10$&$0.23\pm0.06$& 0.19 &0.18& $0.14\pm0.05$ \\
  $D^0\to \eta'\rho^0$ & $C_P$,$C_V$,($E_P$,$E_V$) &0.10&$0.14\pm0.02$&$0.08\pm0.02$& 0.21 &0.27&  \\
  $D^0\to \eta'\omega$ & $C_P$,$C_V$,($E_P$,$E_V$) &0.001&$0.07\pm0.02$&$0.0001\pm0.0001$& 0.04&0.02&  \\
  $D^+\to \pi^+\rho^0$ & $T_V$,$C_P$,($A_P$,$A_V$) &1.7&&$0.8\pm0.7$& 0.42 &0.58& $0.81\pm0.15$ \\
  $D^+\to \pi^+\omega$ & $T_V$,$C_P$,($A_P$,$A_V$) &0.35&&$0.3\pm0.3$& 0.95 & 0.80 & $<0.34$ \\
  $D^+\to \pi^+\phi$ & $C_P$ &5.9&$6.21\pm0.43$&$5.1\pm1.4$& 5.65 &5.65& $5.42^{+0.22}_{-0.24}$ \\
  $D^+\to \pi^0\rho^+$ & $T_P$,$C_V$,($A_P$,$A_V$) &3.7&&$3.5\pm1.6$& 2.7 &2.5&  \\
  $D^+\to K^+\overline{K^*}^0$ & $T_V$,($A_V$) &2.5&&$4.1\pm1.0$& 3.61 &3.60& $3.675^{+0.14}_{-0.21}$ \\
  $D^+\to \overline{K}^0K^{*+}$ & $T_P$,($A_P$) &1.70&&$12.4\pm2.4$& 11 &11& $32\pm14$ \\
  $D^+\to \eta\rho^+$ & $T_P$,$C_V$,($A_P$,$A_V$) &0.002&&$0.4\pm0.4$& 0.7 &2.2& $<15$ \\
  $D^+\to \eta'\rho^+$ & $T_P$,$C_V$,($A_P$,$A_V$) &1.3&&$0.8\pm0.1$& 0.7 &0.8&  \\
  $D_s^+\to \pi^+K^{*0}$ & $T_V$,($A_V$) &3.3&&$1.5\pm0.7$& 2.52 &2.35& $2.25\pm0.39$ \\
  $D_s^+\to \pi^0K^{*+}$ & $C_V$,($A_V$) &0.29&&$0.1\pm0.1$& 0.8 &1.0&  \\
  $D_s^+\to K^+\rho^0$ & $C_P$,($A_P$) &2.4&&$1.0\pm0.6$& 1.9 &2.5& $2.7\pm0.5$ \\
  $D_s^+\to K^+\omega$ & $C_P$,($A_P$) &0.72&&$1.8\pm0.7$& 0.6 &0.07& $<2.4$ \\
  $D_s^+\to K^+\phi$ & $T_V$,$C_P$,($A_V$) &0.15&&$0.3\pm0.3$& 0.166 &0.166& $0.184\pm0.045$ \\
  $D_s^+\to K^0\rho^+$ & $T_P$,($A_P$) &19.5&&$7.5\pm2.1$& 9.1 &9.6&  \\
  $D_s^+\to \eta K^{*+}$ & $T_P$,$C_V$,($A_P$,$A_V$) &0.24&&$1.0\pm0.4$& 0.2 &0.2&  \\
  $D_s^+\to \eta'K^{*+}$ & $T_P$,$C_V$,($A_P$,$A_V$) &0.24&&$0.6\pm0.2$& 0.2 &0.2 & \\
  \hline\hline
  \end{tabular}
\end{table}

\begin{table}[!htbh]
  \centering
  \caption{Same as Table~\ref{1tb:AbrA} for the doubly Cabibbo-suppressed
  $D\to PV$ decays in units of $10^{-4}$, except with the absence of Br(FSI).}\label{3tb:CbrC}
  \begin{tabular}[t]{cccccccc}\hline\hline
  Modes & Amplitudes & Br(diagrammatic) &Br(pole) & Br(FAT)&Br(FAT[mix]) & Br(exp)\\\hline
  $D^0\to \pi^0K^{*0}$ & $C_P$,($E_V$) &$0.54\pm0.18$&$0.8\pm0.3$& 1.0 &0.9&  \\
  $D^0\to \pi^-K^{*+}$ & $T_P$,($E_V$) &$3.59\pm0.72$&$2.7\pm0.6$& 4.82 &4.72& $3.39\pm1.41$ \\
  $D^0\to K^+\rho^-$ & $T_V$,($E_P$) &$1.45\pm0.17$&$0.9\pm0.3$& 1.4 & 1.5& \\
  $D^0\to K^0\rho^0$ & $C_V$,($E_P$) &$0.91\pm0.51$&$0.5\pm0.2$& 0.4 &0.3& \\
  $D^0\to K^0\omega$ & $C_V$,($E_P$) &$0.58\pm0.40$&$0.7\pm0.2$& 0.6 &0.6 & \\
  $D^0\to K^0\phi$ & $E_V$ &$0.06\pm0.05$&$0.20\pm0.06$& 0.2 &0.2 & \\
  $D^0\to \eta K^{*0}$ & $C_P$,($E_P$,$E_V$) &$0.33$&$0.08$& 0.2 &0.2 & \\
  $D^0\to \eta'K^{*0}$ & $C_P$,($E_P$,$E_V$) &$0.0040\pm0.0006$&$0.004\pm0.001$& 0.005 &0.005 & \\
  $D^+\to \pi^+K^{*0}$ & $C_P$,($A_V$) &&$2.2\pm0.9$& 3.33 &3.33& $3.75\pm0.60$ \\
  $D^+\to \pi^0K^{*+}$ & $T_P$,($A_V$) &&$4.0\pm0.9$& 4.0 & 3.9& \\
  $D^+\to K^+\rho^0$ & $T_V$,($A_P$) &&$0.5\pm0.4$& 1.9 &2.4 &$2.0\pm0.5$ \\
  $D^+\to K^+\omega$ & $T_V$,($A_P$) &&$1.8\pm0.5$& 0.9 &0.7 & \\
  $D^+\to K^+\phi$ & $A_V$ &&$0.2\pm0.2$& 0.01 & 0.02& \\
  $D^+\to K^0\rho^+$ & $C_V$,($A_P$) &&$0.5\pm0.4$& 2.3 & 3.3& \\
  $D^+\to \eta K^{*+}$ & $T_P$,($A_P$,$A_V$) &&$1.4\pm0.2$& 1.0 & 1.0& \\
  $D^+\to \eta'K^{*+}$ & $T_P$,($A_P$,$A_V$) &&$0.020\pm0.007$& 0.01 &0.01 & \\
  $D_s^+\to K^+K^{*0}$ & $T_V$,$C_P$ &$0.20\pm0.05$&$0.2\pm0.2$& 0.23 &0.23& $0.90\pm0.53$ \\
  $D_s^+\to K^0K^{*+}$ & $T_P$,$C_V$ &$1.17\pm0.86$&$2.3\pm0.6$& 1.2 & 1.1 & \\
  \hline\hline
  \end{tabular}
\end{table}

\section{Direct $CP$ Asymmetries}
In this section we predict the direct $CP$ asymmetries of
the $D\to PV$ decays, which is defined are
\be
A_{CP}=\frac{\Gamma(D\to PV)-\Gamma(\overline{D}\to \overline{P}\overline{V})}
{\Gamma(D\to PV)+\Gamma(\overline{D}\to \overline{P}\overline{V})},
\ed
by estimating the penguin contributions in the FAT approach.
The quark-loop and magnetic penguin contributions are included and
absorbed into the Wilson coefficients of the
penguin operators. It has been found that the strong phases from
the quark loops and from the scalar penguin annihilation dominate the
direct $CP$ asymmetries \cite{Li:2012cfa}.

\subsection{Parametrization of penguin amplitudes}
The effective weak Hamiltonian for the penguin contributions is written as
\begin{equation}
\Delta H_{eff}=-\frac{G_F}{\sqrt{2}}V^*_{cb}V_{ub}\left[\sum_{i=3}^6
C_i(\mu)O_i(\mu)+C_{8g}(\mu)O_{8g}(\mu)\right],
\end{equation}
where the QCD-penguin and chromomagnetic-penguin operators are defined by
\begin{equation}\begin{split}
&O_3=\Sigma_q(\overline{u}_{\alpha}c_{\alpha})_{V-A}(\overline{q}_{\beta}q_{\beta})_{V-A},\\
&O_4=\Sigma_q(\overline{u}_{\alpha}c_{\beta})_{V-A}(\overline{q}_{\beta}q_{\alpha})_{V-A},\\
&O_5=\Sigma_q(\overline{u}_{\alpha}c_{\alpha})_{V-A}(\overline{q}_{\beta}q_{\beta})_{V+A},\\
&O_6=\Sigma_q(\overline{u}_{\alpha}c_{\beta})_{V-A}(\overline{q}_{\beta}q_{\alpha})_{V+A},\\
&O_{8g}=\frac{g_s}{8\pi^2}m_c\overline{u}\sigma_{\mu\nu}(1+\gamma_5)T^aG^{a\mu\nu}c,
\end{split}\end{equation}
with $m_c$ being the charm quark mass, $T^a$ being a color matrix, and
$G^{a\mu\nu}$ being the gluon field tensor. The eight topological penguin diagrams
for the $D\to PV$ decays are displayed in Fig.~2, in
which the color-favored penguin amplitude $PT_{P(V)}$, the color-suppressed
penguin amplitude $PC_{P(V)}$, the gluon-annihilation penguin amplitude
$PE_{P(V)}$, and the gluon-exchange penguin amplitude $PA_{P(V)}$ correspond
to the tree amplitudes $T_{P(V)}$, $C_{P(V)}$, $E_{P(V)}$, and $A_{P(V)}$,
respectively.

\begin{figure}[!ht]
  \begin{center}
  \includegraphics[scale=0.5]{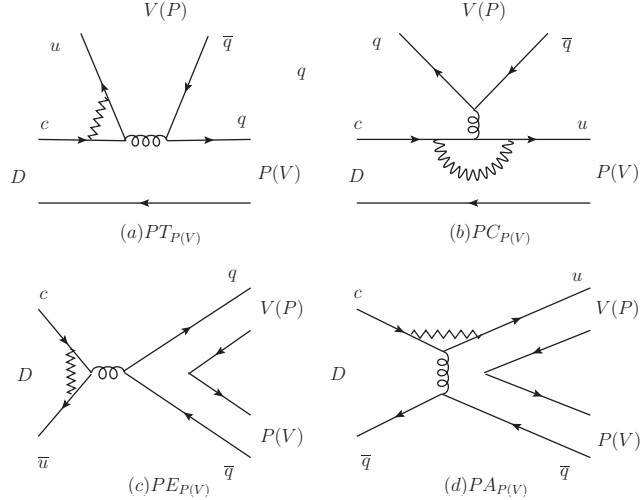}
  \vspace{-7cm}
  \caption{Topological penguin diagrams contributing to the $D\to PV$ decays with
  (a) the color-favored penguin amplitude $PT_{P(V)}$, (b) the color-suppressed
  penguin amplitude $PC_{P(V)}$, (c) the gluon-annihilation penguin amplitude
  $PE_{P(V)}$, and (d) the gluon-exchange penguin amplitude $PA_{P(V)}$.}
  \end{center}\label{Fig:II}
\end{figure}

The contributions from the $(V-A)(V-A)$ operators $O_{3,4}$
can be simply obtained by substituting the associated Wilson
coefficients and CKM matrix elements  in the tree amplitudes,
while the contributions from the $(V-A)(V+A)$ operators $O_{5,6}$ need to be
treated separately. The nonfactorizable contributions to the color-favored
penguin amplitudes are ignored as in the color-favored tree amplitudes.
Since a vector meson cannot be generated from the scalar or pseudoscalar
operator, $PT_P$ does not receive contributions from $O_5$ or $O_6$.
The penguin amplitude $PT_V$ is expressed as
\begin{equation}\begin{split}\label{ptv}
PT_V&=-{G_F\over\sqrt{2}}V^*_{cb}V_{ub}
[a_4(\mu)\langle V|(\overline{q}c)_{V-A}|D\rangle\langle P|(\overline{u}q)_{V-A}|0\rangle
-2a_6(\mu)\langle V|(\overline{q}c)_{S-P}|D\rangle\langle P|(\overline{u}q)_{S+P}|0\rangle]\\
&=-{G_F\over\sqrt{2}}V^*_{cb}V_{ub}[a_4(\mu)-r_Xa_6(\mu)]
f_Pm_VA_0^{DV}(m_P^2)2(\varepsilon\cdot p_D),
\end{split}\end{equation}
with the chiral factor $r_X=2m^P_0/m_c$ and the Wilson coefficients
$a_4=C_4+C_3/N_c$ and $a_6=C_6+C_5/N_c$.

A remark is in order. The penguin operators, with a sum over light
quark flavors, form a U-spin singlet, but the tree operators do not.
It is then expected from symmetry considerations that the penguin matrix
elements may have magnitudes and strong phases different from those of the
tree ones. Take the $D_s^+\to\pi^+K^{*0}$ channel as an example.
The ${\bar s}s$ quark pair in $O_{3,4}$ can also contribute to this decay through
final-state rescattering $\bar{s}s\to\bar{d}d$, that then introduces
an additional source of strong phases and differentiates the $O_{3,4}$
amplitudes from the $O_{1,2}$ amplitudes. However, our formalism relies on
the factorization of short-distance and long-distance dynamics, so the weak
vertex is regarded as a hard vertex. The $s$ and $\bar{s}$ quarks emitted
from the penguin operator fly back to back, and the chance for them to have
rescattering is small; namely, final-state rescattering is regarded as a
subleading effect. We then have specific quark flavors for the external
lines of the decay, to which the tree operators
$O_1=(\bar{u}_\alpha d_\beta)_{V-A}(\bar{d}_\beta c_\alpha)_{V-A}$
and $O_2=(\bar{u}_\alpha d_\alpha)_{V-A}(\bar{d}_\beta c_\beta)_{V-A}$, and
only the $\bar d d$ components of the penguin operators,
$O_3=(\bar{u}_\alpha c_\alpha)_{V-A}(\bar{d}_\beta d_\beta)_{V-A}$
and $O_4=(\bar{u}_\alpha c_\beta)_{V-A}(\bar{d}_\beta d_\alpha)_{V-A}$, contribute
(not considering $O_{5,6}$ here). The above $O_{1(2)}$ and $O_{3(4)}$
are identical according to the Fiertz identity, and they lead to the same hadronic
matrix elements.

The factorizable contributions to $PC_{P,V}$ from $O_{5,6}$ are
easily derived with the relations between the $(V+A)$ and
$(V-A)$ currents used, $\langle V|(\overline{q}_1q_2)_{V+A}|0\rangle=
\langle V|(\overline{q}_1q_2)_{V-A}|0\rangle$ and
$\langle P|(\overline{q}_1q_2)_{V+A}|0\rangle=
-\langle P|(\overline{q}_1q_2)_{V-A}|0\rangle$. The nonfactorizable
contributions from $O_4$ and $O_6$ are related to the tree
contributions in the following way.
Since the hadronic matrix element
of the $(V-A)(V-A)$ operator has been parametrized as
the product of $(1/N_c+\chi^C e^{i\phi^C})$, the decay
constant, and the form factor as shown in Eqs.~(\ref{fdp})-(\ref{a12}),
the nonfactorizable contributions from $O_2$
and $O_4$ carry the same strong phase $\phi^C$.
It has been confirmed by PQCD analytical formulas
for two-body hadronic $D$ meson decays \cite{ZLL13} that
the nonfactorizable contributions from $O_4$ and $O_6$ to $PC_V$ are identical
without including the Wilson coefficients $C_4(\mu)$ and $C_6(\mu)$.
Relative to $PC_V$, an additional negative
sign is added to the contribution from $O_6$ to $PC_P$.
We then arrive at the parametrization
of the color-suppressed penguin amplitudes
\begin{equation}\begin{split}\label{pc}
&PC_P=-{G_F\over\sqrt{2}}V^*_{cb}V_{ub}\left[a_3^P(\mu)+a_5^P(\mu)\right]
f_Vm_VF_1^{DP}(m_V^2)2(\varepsilon\cdot p_D),\\
&PC_V=-{G_F\over\sqrt{2}}V^*_{cb}V_{ub}\left[a_3^V(\mu)-a_5^V(\mu)\right]
f_Pm_VA_0^{DV}(m_P^2)2(\varepsilon\cdot p_D),
\end{split}\end{equation}
with the Wilson coefficients
\begin{eqnarray}
a_3^{P(V)}(\mu)&=&C_3(\mu)+C_4(\mu)\left({1\over N_c}
+\chi^C_{P(V)}e^{i\phi^C_{P(V)}}\right),\nonumber\\
a_5^{P(V)}(\mu)&=&C_5(\mu)+C_6(\mu)\left({1\over N_c}-\chi^C_{P(V)}e^{i\phi^C_{P(V)}}\right).
\end{eqnarray}

All the factorizable contributions to the annihilation-type penguin diagrams
are neglected because of the helicity suppression, except those to the
diagrams $PA_{P,V}$ from $O_{5,6}$. They are expressed as
\begin{equation}\begin{split}
PA_{P(V)}^{f}=&-\frac{G_F}{\sqrt{2}}V^*_{cb}V_{ub}a_6(\mu)
\langle VP(PV)|(\overline{u}c)_{V-A}(\overline{q}q)_{V+A}|D\rangle\\
=&-\frac{G_F}{\sqrt{2}}V^*_{cb}V_{ub}a_6(\mu)(-2)
\langle VP(PV)|(\overline{u}q)_{S+P}|0\rangle\langle 0|(\overline{q}c)_{S-P}|D\rangle,
\label{papv}
\end{split}\end{equation}
after the Fierz transformation and the factorization hypothesis are applied.
In the pole resonance model, Eq.~(\ref{papv}) becomes
\begin{equation}\begin{split}\label{pole}
PA_P^{f}&=-\frac{G_F}{\sqrt{2}}V^*_{cb}V_{ub}a_6(\mu)
(-2)\langle VP|H_s|P^*\rangle\frac{1}{m_D^2-m_{P^*}^2}
\langle P^*|(\overline{u}q)_{S+P}|0\rangle \langle0|(\overline{q}c)_{S-P}|D\rangle\\
&=2\frac{G_F}{\sqrt{2}}V^*_{cb}V_{ub}a_6(\mu)
(2g_{PPV} p_D\cdot\epsilon_V)\frac{1}{m_D^2-m_{P^*}^2}
(f_{P^*}m^0_{P^*})(f_D\frac{m_D^2}{m_c}),\\
PA_V^{f}&=-\frac{G_F}{\sqrt{2}}V^*_{cb}V_{ub}a_6(\mu)
(-2)\langle PV|H_s|P^*\rangle \frac{1}{m_D^2-m_{P^*}^2}
\langle P^*|(\overline{u}q)_{S+P}|0\rangle \langle 0|(\overline{q}c)_{S-P}|D\rangle\\
&=2\frac{G_F}{\sqrt{2}}V^*_{cb}V_{ub}a_6(\mu)
(-2g_{PPV}p_D \cdot\epsilon_V)\frac{1}{m_D^2-m_{P^*}^2}
(f_{P^*}m^0_{P^*})(f_D\frac{m_D^2}{m_c}),
\end{split}\end{equation}
where $P^*$ represents the pole resonant pseudoscalar meson and $H_s$
is the corresponding strong Hamiltonian. The corresponding effective
coupling constants $g_{PPV}$'s are obtained from $\rho\to\pi\pi$,
$K^*(892)^0\to\pi^+K^-$, and $\phi\to K^+K^-$, as handled in Ref.
\cite{Fusheng:2011tw}. We set $g_{PPV}$ to be $g_q=4.2$ if none of the
three strongly coupled mesons contains $s$ quarks, to be $g_s=4.6$ if
two of them contain $s$ quarks, and to be $g_{ss}=4.5$ if all of them
contain $s$ quarks.

The PQCD approach \cite{ZLL13} suggests that the nonfactorizable
contributions to $PA_{P,V}$ from $O_5$ almost vanish, leading to the parametrization
\begin{equation}\begin{split}
&PA_P^{nf}\propto C_3(\mu)\chi^A_{q(s)}e^{i\phi^A_{q(s)}},\\
&PA_V^{nf}\propto C_3(\mu)\chi^A_{q(s)}e^{i\phi^A_{q(s)}}.\label{paf}
\end{split}\end{equation}
The sum of Eqs.~(\ref{pole}) and (\ref{paf}) completes the parametrization
of the amplitudes $PA_{P,V}$, which turn out to carry strong phases different
from those of the tree amplitudes $A_{P,V}$ in Eq.~(\ref{eq:A}).
For the nonfactorizable contributions to the
amplitudes $PE_{P,V}$, the PQCD approach \cite{ZLL13} suggests that the
dominant pieces from $O_4$ and $O_6$ are formulated in the same way as
\begin{equation}\begin{split}
&PE_P\propto[C_4(\mu)-C_6(\mu)]\chi^E_{q(s)}e^{i\phi^E_{q(s)}},\\
&PE_V\propto[C_4(\mu)-C_6(\mu)]\chi^E_{q(s)}e^{i\phi^E_{q(s)}}.\label{panf}
\end{split}\end{equation}

The quark-loop contributions from the tree operators
can be absorbed into the Wilson coefficients as \cite{BBNS99}
\begin{equation}\begin{split}
&C_{3,5}(\mu)\to C_{3,5}-\frac{\alpha_s(\mu)}{8\pi N_c}
\sum_{q=d,s}{\lambda_q\over\lambda_b}C^q(\mu,\langle l^2\rangle),\\
&C_{4,6}(\mu)\to C_{4,6}+\frac{\alpha_s(\mu)}{8\pi}
\sum_{q=d,s}{\lambda_q\over\lambda_b}C^q(\mu,\langle l^2\rangle),
\end{split}\end{equation}
where $\langle l^2\rangle$ is the averaged invariant mass squared
of the virtual gluon emitted from the quark loop; $\lambda_q$
is defined as $V^*_{cq}V_{uq}$ for the
quark $q=d$, $s$ or $b$; and the function $C^q$ is given by
\begin{equation}
C^q(\mu,\langle l^2\rangle)=\left[-\frac{2}{3}-4\int_0^1dxx(1-x)\ln\frac{m_q^2-x(1-x)
\langle l^2\rangle}{\mu^2}\right]C_2(\mu),
\end{equation}
with the quark mass $m_q$.
We set the value of $\langle l^2\rangle$ to be $(P_P/2+P_V/2)^2=m_D^2/4$ by
assuming that each spectator of a light meson is likely to carry half of the
meson momentum. We have checked that our predictions for direct $CP$
asymmetries stayed stable as $\langle l^2\rangle$ ranges from $m_D^2/25$ to $m_D^2$.
The chromomagnetic-penguin contribution can be further absorbed
into the Wilson coefficients, leading to \cite{BBNS99}
\begin{equation}\begin{split}
C_{3,5}(\mu)&\to C_{3,5}-\frac{\alpha_s(\mu)}{8\pi N_c}
\sum_{q=d,s}{\lambda_q\over\lambda_b}C^q(\mu,\langle l^2\rangle)
+\frac{1}{N_c}\frac{\alpha_s(\mu)}{4\pi}\frac{m_c^2}{\langle l^2\rangle}
[C_{8g}(\mu)+C_5(\mu)],\\
C_{4,6}(\mu)&\to C_{4,6}+\frac{\alpha_s(\mu)}{8\pi }
\sum_{q=d,s}{\lambda_q\over\lambda_b}C^q(\mu,\langle l^2\rangle)
-\frac{\alpha_s(\mu)}{4\pi}\frac{m_c^2}{\langle l^2\rangle}[C_{8g}(\mu)+C_5(\mu)].
\end{split}\end{equation}

\subsection{Penguin-induced $CP$ violation}

We list the predicted direct $CP$ asymmetries in the $D\to PV$
decays without and with various corrections (QCD-penguins,
chromomagnetic penguins, quark loops, pole resonances, and
$\rho-\omega$ mixing) in Table~\ref{cpa1}. For the $D^0$ decays,
the direct $CP$ asymmetries cannot be measured directly in
experiments owing to the $D^0$-$\overline{D}^0$ mixing. However,
we can obtain the time-integrated $CP$ asymmetries by adding the
contributions from the indirect $CP$ asymmetries to the direct
ones, as done in Ref. \cite{arXiv:1112.0938}. It can be found
from Table~\ref{cpa1} that the $D^0\to K^0\overline{K}^{*0}$ and
$D^0\to \overline{K}^0K^{*0}$ modes do not receive contributions
from the quark loops or the chromomagnetic penguins, since these
two contributions to the Wilson coefficients $C_4(\mu)$ and
$C_6(\mu)$ cancel exactly with each other in the amplitudes
$PE_{P,V}$. We can also find that the direct $CP$ asymmetries of
the $\omega$-involved modes change considerably with the
$\rho$-$\omega$ mixing effect. Similarly to the case of
branching ratios, the mixing effect lowers the $\omega$ meson
decay constant, which has considerable influence on both the
tree and penguin amplitudes of the $\omega$-involved modes.

\begin{table}[!htbh]
  \centering
  \caption{Direct $CP$ asymmetries for the $D\to PV$ decays in units of $10^{-3}$.
  The results excluding and including various corrections (QCD-penguins,
  chromomagnetic penguins, quark loops, pole resonances, and $\rho-\omega$ mixing)
  one by one are listed. The relevant amplitudes of the decays are also shown,
  with those outside the parentheses being dominant.}\label{cpa1}
  \begin{tabular}[t]{ccccccc}\hline\hline
  Modes & Amplitudes & $A_{CP}$(tree) & $A_{CP}$(+penguin) & $A_{CP}$(+cm,ql) & $A_{CP}$(+pole) & $A_{CP}$(mixing) \\\hline
  $D^0$$\to$$\pi^+$$\rho^-$ & $PT$, $PA$, ($PE$) & 0 & -0.03 & 0.02 & -0.02& -0.03\\
  $D^0$$\to$$\pi^0$$\rho^0$ & $PT$, $PC$, ($PE$, $PA$) & 0 & -0.01 & -0.02 & -0.02& -0.03\\
  $D^0$$\to$$\pi^0$$\omega$ & $PT$, $PC$, ($PE$, $PA$) & 0 & 0.0002 & 0.04 & 0.04& 0.02\\
  $D^0$$\to$$\pi^0$$\phi$ & $PC$ & 0 & -0.0002 & -0.0002 & -0.0002& -0.0002\\
  $D^0$$\to$$\pi^-$$\rho^+$ & $PT$, $PA$, ($PE$) & 0 & 0.01 & -0.04 & -0.01& -0.01\\
  $D^0$$\to$$K^+$$K^{*-}$ & $PT$, $PA$, ($PE$) & 0 & 0.05 & -0.03 & -0.01& -0.01\\
  $D^0$$\to$$K^0$$\overline{K}^{*0}$ & $PE$ & -0.7 & -0.7 & -0.7 & -0.7& -0.7\\
  $D^0$$\to$$\overline{K}^0$$K^{*0}$ & $PE$ & -0.7 & -0.7 & -0.7 & -0.7& -0.7\\
  $D^0$$\to$$K^-$$K^{*+}$ & $PT$, $PA$, ($PE$) & 0 & -0.04 & 0.03 & 0& 0\\
  $D^0$$\to$$\eta$$\rho^0$ & $PT$, $PC$, ($PE$, $PA$) & 0.8 & 0.8 & 0.8 & 0.8& 1.0\\
  $D^0$$\to$$\eta$$\omega$ & $PT$, $PC$, ($PE$, $PA$) & -0.2 & -0.1 & -0.2 & -0.2& -0.1\\
  $D^0$$\to$$\eta$$\phi$ & $PC$, ($PE$) & 0 & 0.003 & 0.003 & 0.003 & 0.003\\
  $D^0$$\to$$\eta'$$\rho^0$ & $PT$, $PC$, ($PE$, $PA$) & -0.5 & -0.3 & -0.2 & -0.2& -0.1\\
  $D^0$$\to$$\eta'$$\omega$ & $PT$, $PC$, ($PE$, $PA$) & 1.8 & 1.2 & 1.2 & 1.2& 2.2\\
  $D^+$$\to$$\pi^+$$\rho^0$ & $PT$, $PC$, $PA$ & 0 & -0.5 & -0.5 & 0.7& 0.5\\
  $D^+$$\to$$\pi^+$$\omega$ & $PT$, $PC$, ($PA$) & 0 & 0.06 & 0.03 & 0.03& -0.05\\
  $D^+$$\to$$\pi^+$$\phi$ & $PC$ & 0 & -0.0001 & -0.0001 & -0.0001& -0.0001\\
  $D^+$$\to$$\pi^0$$\rho^+$ & $PT$, $PC$, $PA$  & 0 & 0.03 & -0.2 & 0.2& 0.2\\
  $D^+$$\to$$K^+$$\overline{K}^{*0}$ & $PT$, $PA$  & 0.1 & 0.5 & 0.1 & 0.2& 0.2\\
  $D^+$$\to$$\overline{K}^0$$K^{*+}$ & $PT$, $PA$  & 0.08 & 0.07 & 0.15 & 0.04& 0.04\\
  $D^+$$\to$$\eta$$\rho^+$ & $PT$, $PC$, ($PA$) & -0.7 & -0.7 & -0.7 & -0.7& -0.6\\
  $D^+$$\to$$\eta'$$\rho^+$ & $PT$, $PC$, ($PA$) & 0.2 & 0.1 & 0.3 & 0.3& 0.5\\
  $D_s^+$$\to$$\pi^+$$K^{*0}$ & $PT$, $PA$ & 0.2 & 0.2 & 0.2 & -0.2& -0.1\\
  $D_s^+$$\to$$\pi^0$$K^{*+}$ & $PT$, $PC$, $PA$ & 0.2 & 0.2 & 0.3 & -0.3& -0.2\\
  $D_s^+$$\to$$K^+$$\rho^0$ & $PT$, $PC$, $PA$ & -0.01 & -0.05 & -0.1 & 0.3& 0.3\\
  $D_s^+$$\to$$K^+$$\omega$ & $PT$, $PC$, $PA$ & 0.03 & 0.09 & 0.2 & -0.6& -2.3\\
  $D_s^+$$\to$$K^+$$\phi$ & $PT$, $PC$, $PA$ & 0 & 0.4 & 0.3 & -0.8& -0.8\\
  $D_s^+$$\to$$K^0$$\rho^+$ & $PT$, $PA$ & 0.04 & 0.03 & -0.02 & 0.2& 0.3\\
  $D_s^+$$\to$$\eta$$K^{*+}$ & $PT$, $PC$, $PA$ & 0.5 & 0.4 & 0.8 & -0.3& 1.1\\
  $D_s^+$$\to$$\eta'$$K^{*+}$ & $PT$, $PC$, $PA$ & -0.1 & -0.1 & -0.2 & -0.4& -0.5\\
  \hline\hline
  \end{tabular}
\end{table}

The $D\to PP$ analysis has indicated that the direct $CP$ asymmetries of
the $D^0\to\pi^+\pi^-$ and $K^+K^-$ decays reach ${\cal O}(10^{-4})$
\cite{Li:2012cfa}. It seems that the direct $CP$ asymmetries of the
corresponding $D\to PV$ decays, such as $D^0\to \pi^+\rho^-$,
$\pi^-\rho^+$, $K^+K^{*-}$, and $K^-K^{*+}$, should be of the same order.
However, tiny values for these four modes are predicted as shown in
Table~\ref{cpa1}. We investigate the $D^0\to \pi^+\rho^-$ decay
specifically, for which the direct $CP$ asymmetry receives contributions
mainly from the penguin amplitudes $PT_{V}$ and $PA_{V}$ (for which the
nonfactorizable contributions are negligible). According to
Eqs.~(\ref{ptv}) and (\ref{pole}), $PT_{V}$ and $PA_{V}$ carry nearly the
same magnitude and phase, but with an opposite sign between them (the
corresponding two amplitudes in $D^0\to\pi^+\pi^-$ have the same sign).
Therefore, it is numerically coincident that they cancel each other,
$PA_{V}+PT_{V}\approx 0$. The small direct $CP$ asymmetries in these
decays are then understood.

The direct $CP$ asymmetries in several modes, including
$D^0\to K^0\overline{K}^{*0}$, $\overline{K}^0K^{*0}$, $\eta\rho^0$, $\eta'\omega$,
$D^+\to\pi^+\rho^0$, $\eta\rho^+$, and $D_s^+\to K^+\omega$, $K^+\phi$,
$\eta K^{*+}$, reach ${\cal O}(10^{-3})$ as shown by Table~\ref{cpa1},
which are expected to be observed at LHCb or Belle II
in the future. In particular, the detecting efficiency of the final states
in the $D^+\to\pi^+\rho^0$ and $D_s^+\to K^+\omega$, $K^+\phi$
decays is high. The direct $CP$ asymmetry in the $D^+\to\pi^+\phi$
mode has been recently measured by LHCb, and the datum
$(-0.04\pm0.14\pm0.13)\%$ \cite{LHCbCPA} is consistent with zero
as predicted in the FAT approach.

The contributions from new physics to
electroweak interactions can be easily absorbed into the Wilson
coefficients in the FAT approach. Given a new-physics model, we can calculate
how the Wilson coefficients are modified in order to match the observed
direct $CP$ asymmetries and then use the new Wilson coefficients to predict
direct $CP$ asymmetries in other modes. For example, if a
new-physics model has a considerable impact
only on the chromomagnetic penguin operator $O_{8g}$, which is
allowed by the constraints from the $D^0$-$\overline{D}^0$ mixing
\cite{ddbar}, we extract $C_{8g}\approx 11$ from the first
measurement of $\Delta A_{\rm CP}$ by LHCb \cite{arXiv:1112.0938}.
Then we predict the direct $CP$ asymmetries in the $D\to PV$
modes and find that two of them are hopefully measured:
about $1\%$ for $D^+\to \pi^+\rho^0$ and about $-1\%$
for $D_s^+\to K^+\phi$. For those decays for which the
tree amplitudes do not contribute to the $CP$ asymmetries, their $CP$
asymmetries are proportional to the penguin amplitudes
\cite{Buccella:1995fsi}. In other words, they are simply proportional
to the QCD-penguin Wilson coefficients $C_{3-6}(\mu)$.
In some new physics models, these coefficients are synchronously
varied and will become about 1 order larger in order to
accommodate the measured $\Delta A_{\rm CP}$. As a consequence, the direct
$CP$ asymmetries of most modes listed in Table~\ref{cpa1}
will be enhanced by 1 order of magnitude.
Specifically, the direct $CP$ asymmetry of the $D^+\to \pi^+\rho^0$
decay can reach 1\% level.

As shown at the end of Sec.~II, the neglected contributions, such
as the nonfactorizable $T$ and the factorizable $E$ and $A$, lead to
small corrections to the branching ratios. The corrections
from the corresponding penguin contributions, parametrized
in a similar way, then modify the predicted direct $CP$ asymmetries. Their
effects can be used to estimate the uncertainties for predictions
in our approach, which are found to be about $17\%$. Besides, the
signs of the predicted direct $CP$ asymmetries never flip.
This level of precision should be acceptable, considering the
tremendous difficulty to analyze $D$ meson decays theoretically.

Finally, the $CP$ asymmetry observables of some neutral $D$ meson
decays with the final states $f$, which follow the definitions in
Ref. \cite{neutralCPA}, are listed in Table~\ref{cpa5}. The
$\rho$-$\omega$ mixing has a negligible influence on these
observables. The other neutral $D$ meson decays are not
considered, since their time evolution effect is tiny.

\begin{table}[!htbh]
  \centering
  \caption{$CP$ asymmetry observables of some neutral $D$ meson decays.}\label{cpa5}
  \begin{tabular}[t]{cccccc}\hline\hline
  Modes & $C_f$ & $S_f$ & $S_{\overline{f}}$ & $D_f$ & $D_{\overline{f}}$  \\\hline
  $D^0$$\to$$\pi^+$$\rho^-$  & -0.4 & -0.1 & -0.1 & 0.9 & 0.9\\
  $D^0$$\to$$\pi^-$$\rho^+$   & 0.4 & 0.1 & 0.1 & 0.9 & 0.9\\
  $D^0$$\to$$K^+$$K^{*-}$  & -0.4 & -0.2 & -0.2 & 0.9 & 0.9\\
  $D^0$$\to$$K^-$$K^{*+}$   & 0.4 & 0.2 & 0.2 & 0.9 & 0.9\\
  \hline\hline
  \end{tabular}
\end{table}

\section{Summary}

In this paper we have analyzed the branching ratios and direct $CP$
asymmetries of the $D\to PV$ decays in the FAT approach, which was
proposed in Ref. \cite{Li:2012cfa}. Briefly speaking, we have
improved the topology parametrization by taking into account
mode-dependent QCD dynamics, for instance, the evolution of the
Wilson coefficients with the energy release in individual modes,
flavor SU(3) symmetry breaking effects, and strong phases from FSI
and from the Glauber gluons in nonfactorizable annihilation-type
amplitudes. The $\rho$-$\omega$ mixing effect has been included,
which improves the global fit to the branching ratios involving the
$\rho^0$ and $\omega$ mesons. The puzzle from the
$D_s^+\to \pi^+\rho^0$, $\pi^+\omega$ branching ratios observed in
the previous studies has been also resolved. Combining the
short-distance dynamics associated with the penguin operators and
the hadronic parameters determined from the global fit to the
measured branching ratios, we have predicted the direct $CP$
asymmetries in the $D\to PV$ decays. The parametrization of some
nonfactorizable contributions from the operator $O_6$ was guided by
the PQCD analysis for two-body hadronic $D$ meson decays.
Fortunately, these contributions do not dominate our predictions
for the direct $CP$ asymmetries in most of the $D\to PV$ modes. It
was found that the direct $CP$ asymmetries in the
$D^0\to K^0\overline{K}^{*0},~\overline{K}^0K^{*0}$,
$D^+\to\pi^+\rho^0$, and $D_s^+\to K^+\omega,~K^+\phi$ decays reach
${\cal O}(10^{-3})$, which may be observed at the LHCb or Belle II
experiment. The $CP$ asymmetry observables of some neutral $D$ meson
decays have also been calculated. Many of our predictions can be
confronted with future data.

\section{Acknowledgements}

We are grateful to Zhi-Tian Zou and Pei-Lian Liu for useful discussions.
The work was partly supported by the National Science Council of R.O.C.
under the Grant No. NSC-101-2112-M-001-006-MY3 and National Science Foundation
of China under the Grants No.11375208, No. 11228512, and No. 11235005.

\begin{appendix}
\section{STRONG MATRIX ELEMENTS}

In this appendix we determine the relative sign between the hadronic matrix elements
$\langle PV|H_s|P^*\rangle$ (with the pseudoscalar meson $P$ being emitted) and
$\langle VP|H_s|P^*\rangle$ (with the vector meson $V$ being emitted)
in Eq.~(\ref{pole}). Since the strong vertex $H_s\propto
iV_{\mu}(P_1\partial^\mu P_2-P_2\partial^\mu P_1)$ is antisymmetric under the
exchange of the mesons $P_1$ and $P_2$, we need to differentiate $P_1$ and $P_2$ to
avoid a wrong sign. It is achieved by comparing an emission amplitude in the pole
resonance model to that in the naive factorization method. We use the emission
amplitude for the decay $D^0\to\rho^+\pi^-$ to fix the sign of
$\langle VP|H_s|P^*\rangle$
and use that of the decay $D^0\to\pi^+\rho^-$ to fix the sign of
$\langle PV|H_s|P^*\rangle$.
We consider the following decay amplitudes:
\begin{equation}\begin{split}
\langle \rho^+\pi^-|H_{eff}|D^0\rangle&=\langle\rho^+|(\overline{u}d)_{V-A}|0\rangle
\langle 0|(\overline{d}c)_{V-A}|D^{*+}\rangle\frac{1}{m_\rho^2-m_{D^*}^2}
\langle D^{*+}\pi^-|H_s|D^0\rangle\\
&=f_\rho m_\rho f_{D^*}m_{D^*}\frac{1}{m_{D^*}^2-m_\rho^2}\langle D^{*+}\pi^-|H_s|D^0\rangle,\\
\langle \pi^+\rho^-|H_{eff}|D^0\rangle&=\langle \pi^+|(\overline{u}d)_{V-A}|0\rangle
\langle 0|(\overline{d}c)_{V-A}|D^+\rangle\frac{1}{m_\pi^2-m_B^2}
\langle D^+\rho^-|H_s|D^0\rangle\\
&=-f_\pi f_Dm_\pi^2\frac{1}{m_D^2-m_\pi^2}\langle D^+\rho^-|H_s|D^0\rangle.\\
\end{split}\end{equation}
For the emission amplitudes to get positive values as in the naive factorization,
$\langle D^{*+}\pi^-|H_s|D^0\rangle$ should be positive, and
$\langle D^+\rho^-|H_s|D^0\rangle$ should be negative. Therefore,
we have the strong matrix elements
\begin{equation}\begin{split}
&\langle VP|H_s|P^*\rangle=2g_{PPV}~p_D\cdot\epsilon_V,\\
&\langle PV|H_s|P^*\rangle =-2g_{PPV}~p_D\cdot\epsilon_V.
\end{split}\end{equation}

\section{$\rho$-$\omega$ Mixing}

In this appendix we formulate the $\rho$-$\omega$ mixing and its effect
on the decay constants of the $\rho^0$ and $\omega$ mesons. As elaborated
in Sec.~II, this mixing plays an important role in the evaluation of
the branching ratio and the direct $CP$ asymmetry of a decay mode involving
$\rho^0$ or $\omega$.
The isospin eigenstates introduced in Eq.~(\ref{eq:mixing}) are written as
\begin{equation}\begin{split}
&|\rho^0_I\rangle={1\over\sqrt{2}}(|\overline{u}u\rangle-|\overline{d}d\rangle),\\
&|\omega_I\rangle={1\over\sqrt{2}}(|\overline{u}u\rangle+|\overline{d}d\rangle).
\end{split}\end{equation}
The decay constants $f_\rho^0$ and $f_\omega^0$ of the isospin eigenstates are defined via
\begin{equation}\begin{split}
&\langle 0|{1\over\sqrt{2}}(\overline{u}\gamma_\mu u-\overline{d}\gamma_\mu d)|\rho^0\rangle
=f_\rho^0 m_\rho\varepsilon^\rho_\mu,\\
&\langle 0|{1\over\sqrt{2}}(\overline{u}\gamma_\mu u+\overline{d}\gamma_\mu d)|\omega\rangle
=f_\omega^0 m_\omega\varepsilon^\omega_\mu,
\end{split}\end{equation}
where $m_{\rho,\omega}$ and $\varepsilon^{\rho,\omega}$ are
the the physical masses and polarization vectors, respectively.

We can obtain the decay constant of a light neutral vector meson $V^0$
through the $V^0\to e^+e^-$ decay width, which occurs through
the electromagnetic current
\begin{equation}\begin{split}
j^{em}_\mu&=Q_u\overline{u}\gamma_\mu u+Q_d\overline{d}\gamma_\mu d\\
&={1\over3\sqrt{2}}j^{I=0}_\mu+{1\over\sqrt{2}}j^{I=1}_\mu,
\end{split}\end{equation}
with the quark charges $Q_{u,d}$ and the isospin currents $j^{I=0,1}_\mu
=(\overline{u}\gamma_\mu u\pm\overline{d}\gamma_\mu d)/\sqrt{2}$.
The $V^0\to e^+e^-$ amplitude is proportional to the matrix
element $\langle 0|j^{em}_\mu|V^0\rangle$, for which we have, from Eq.~(\ref{eq:mixing}),
\begin{equation}\begin{split}
&\langle 0|j^{em}_\mu|\rho^0\rangle=({1\over\sqrt{2}}
f_\rho^0 m_\rho-{\epsilon\over3\sqrt{2}} f_\omega^0m_\omega)\varepsilon^\rho_\mu \equiv\mathcal{T}_\rho\varepsilon^\rho_\mu,\\
&\langle 0|j^{em}_\mu|\omega\rangle=({1\over3\sqrt{2}}
f_\omega^0 m_\omega+{\epsilon\over\sqrt{2}}f_\rho^0m_\rho)\varepsilon^\omega_\mu \equiv\mathcal{T}_\omega\varepsilon^\omega_\mu.
\end{split}\end{equation}
The decay width is then expressed as
\begin{equation}
\Gamma(V^0\to e^+e^-)={4\pi\over3}{\alpha^2\over m_V^3}|\mathcal{T}_V|^2,
\end{equation}
with the fine structure constant $\alpha$.
The decay constant of the physical $\rho^0(\omega)$ meson, $f_{\rho(\omega)}$,
can be read off from the experimental measurements of $\Gamma_{\rho^0(\omega)}$.
Therefore, the physical decay constants are related to those
for the isospin eigenstates via
\begin{equation}\begin{split}
&\left|{1\over\sqrt{2}}f_\rho m_\rho\right|^2=
\left|{1\over\sqrt{2}}f_\rho^0 m_\rho-{\epsilon\over3\sqrt{2}} f_\omega^0 m_\omega\right|^2,\\
&\left|{1\over3\sqrt{2}}f_\omega m_\omega\right|^2=
\left|{1\over3\sqrt{2}}f_\omega^0 m_\omega+{\epsilon\over\sqrt{2}} f_\rho^0 m_\rho\right|^2.
\end{split}\end{equation}
Then, we obtain the decay constants for the isospin eigenstates,
\begin{equation}\begin{split}
&f_\rho^0=f_\rho+{\epsilon\over3}{m_\omega\over m_\rho}f_\omega,\\
&f_\omega^0=f_\omega-3\epsilon{m_\rho\over m_\omega}f_\rho,
\end{split}\end{equation}
where the higher-order terms in $\epsilon$ have been neglected.

\end{appendix}


\begin{thebibliography}{99}

\bibitem{arXiv:1112.0938}
  R.~Aaij {\it et al.}  [LHCb Collaboration],
  Phys. Rev. Lett. {\bf 108}, 111602 (2012);
  M.~Charles [LHCb Collaboration],
  PoS EPS {\bf -HEP2011}, 163 (2011).

\bibitem{CDF12}
T.~Aaltonen et al. [CDF Collaboration], Phys. Rev. Lett. {\bf 109},
111801 (2012).

\bibitem{Belle12}
  B.~R.~Ko [Belle Collaboration],
  PoS ICHEP {\bf 2012}, 353 (2013).



\bibitem{Grossman:2006jg}
  Y.~Grossman, A.~L.~Kagan and Y.~Nir,
  Phys. Rev. D {\bf 75}, 036008 (2007).

\bibitem{Bigi:2011re}
  I.~I.~Bigi, A.~Paul and S.~Recksiegel,
  JHEP {\bf 1106}, 089 (2011).

\bibitem{Brod:2011re}
  J.~Brod, A.~L.~Kagan and J.~Zupan,
  Phys. Rev. D {\bf86}, 014023 (2012).

\bibitem{Cheng:2012wr}
  H.~-Y.~Cheng and C.~-W.~Chiang,
  Phys. Rev. D {\bf 85}, 034036 (2012).

\bibitem{BBNS99}
  M.~Beneke, G.~Buchalla, M.~Neubert and C.~T.~Sachrajda,
  Phys. Rev. Lett.  {\bf 83}, 1914 (1999);
  M.~Beneke, G.~Buchalla, M.~Neubert and C.~T.~Sachrajda,
  Nucl. Phys. {\bf B591}, 313 (2000);
  M.~Beneke and M.~Neubert,
  Nucl. Phys. {\bf B675}, 333 (2003).

\bibitem{BBNS01}
  M.~Beneke, G.~Buchalla, M.~Neubert and C.~T.~Sachrajda,
  Nucl. Phys. {\bf B606}, 245 (2001).

\bibitem{Bhattacharya:2012ah}
  B.~Bhattacharya, M.~Gronau and J.~L.~Rosner,
  Phys. Rev. D {\bf85}, 054014 (2012); Phys. Rev. D {\bf85}, 079901(E) (2012).

\bibitem{Pirtskhalava:2011va}
  D.~Pirtskhalava and P.~Uttayarat,
  Phys.\ Lett.\ B {\bf 712}, 81 (2012)

\bibitem{Hochberg:2011ru}
  Y.~Hochberg and Y.~Nir,
  Phys. Rev. Lett. {\bf108}, 261601 (2012) .

\bibitem{Altmannshofer:2012ur}
  W.~Altmannshofer, R.~Primulando, C.~-T.~Yu and F.~Yu,
  JHEP {\bf1204}, 049 (2012).

\bibitem{Isidori:2011qw}
  G.~Isidori, J.~F.~Kamenik, Z.~Ligeti and G.~Perez,
  Phys. Lett. B {\bf711},  46 (2012).


\bibitem{Feldmann:2012js}
  T.~Feldmann, S.~Nandi and A.~Soni,
  JHEP {\bf1206}, 007 (2012).

\bibitem{Giudice:2012qq}
  G.~F.~Giudice, G.~Isidori and P.~Paradisi,
  JHEP {\bf1204}, 060 (2012).

\bibitem{Chen:2012am}
  C.~-H.~Chen, C.~-Q.~Geng and W.~Wang,
  Phys. Rev. D {\bf85}, 077702 (2012).


\bibitem{Rozanov:2011gj}
  A.~N.~Rozanov and M.~I.~Vysotsky,
  arXiv:1111.6949 [hep-ph].


\bibitem{Li:2012cfa}
  H.~-n.~Li, C.~-D.~Lu and F.~-S.~Yu,
  Phys. Rev. D {\bf 86}, 036012 (2012).

\bibitem{GG89}
  M.~Golden and B.~Grinstein,
  Phys. Lett. B {\bf 222}, 501 (1989).

\bibitem{pietro}
  F.~Buccella, M.~Lusignoli, G.~Miele, A.~Pugliese and P.~Santorelli,
  Phys. Rev. D {\bf 51}, 3478 (1995).

\bibitem{LHCbBB}
R. Aaij et al. [LHCb Collaboration], LHCb-CONF-2013-003.


\bibitem{LHCbB}
R.~Aaij et al. [LHCb Collaboration], 
Phys. Lett. B {\bf723}, 33 (2013).

\bibitem{Cheng:2010ry}
  H.~-Y.~Cheng and C.~-W.~Chiang,
  Phys. Rev. D {\bf 81}, 074021 (2010).

\bibitem{CL11}
  C.~-p.~Chang and H.~-n.~Li,
  Eur. Phys. J. C {\bf 71}, 1687 (2011).

\bibitem{BR10}
  B.~Bhattacharya and J.~L.~Rosner,
  Phys. Rev. D {\bf 81}, 014026 (2010).


\bibitem{Li:2001pqcd}
  Y.~-Y.~Keum, H.~-n.~Li and A.~-I.~Sanda,
  Phys. Lett. B {\bf 504}, 6 (2001);
  C.~-D.~Lu, K.~Ukai, and M.~-Z.~Yang,
  Phys. Rev. D {\bf 63}, 074009 (2001).

\bibitem{Hoang:1995ns}
  N.~L.~Hoang, A.~V.~Nguyen and X.~Y.~Pham,
  Phys.\ Lett.\ B {\bf 357}, 177 (1995).

\bibitem{Li:2011gf}
  H.~-n.~Li and S.~Mishima,
  Phys. Rev. D {\bf 83}, 034023 (2011).

\bibitem{NS08}
  G.~P.~Lepage and S.~J.~Brodsky,
  Phys. Lett. B {\bf 87}, 359 (1979);
  S.~Nussinov and R.~Shrock,
  Phys. Rev. D {\bf 79}, 016005 (2009);
  M.~Duraisamy and A.~L.~Kagan,
  Eur. Phys. J. C {\bf 70}, 921 (2010).

\bibitem{Fusheng:2011tw}
  F.-S.~Yu, X.~-X.~Wang and C.~-D.~Lu,
  Phys. Rev. D {\bf 84}, 074019 (2011).


\bibitem{formfactor}
  D.~Melikhov and B.~Stech,
  Phys. Rev. D {\bf62}, 014006 (2000);
  Y.~L.~Wu, M.~Zhong and Y.~B.~Zuo,
  Int. J. Mod. Phys. A {\bf21}, 6125 (2006).

\bibitem{2012PDG}
  J.~Beringer et~al. [Particle~Data~Group],
  Phys. Rev. D {\bf 86}, 010001 (2012).

\bibitem{Buccella:1995fsi}
  F.~Buccella, M.~Lusignoli, G.~Miele, A.~Pugliese, and P.~Santorelli,
  Phys. Rev. D {\bf 51}, 3478 (1995);
  F.~Buccella, M.~Lusignoli, G.~Mangano, G.~Miele, A.~Pugliese, and P.~Santorelli,
  Phys. Lett. B {\bf 302}, 319 (1993).

\bibitem{CLEO:1998eta'}
  C.~P.~Jessop et~al. [CLEO~Collaboration],
  Phys. Rev. D {\bf 58}, 052002 (1998).

\bibitem{Onyisi:2013bjt}
  P.~U.~E.~Onyisi {\it et al.}  [ CLEO Collaboration],
  Phys.\ Rev.\ D {\bf 88}, 032009 (2013).

\bibitem{Lu:2000hj}
  C.~-D.~Lu and M.~-Z.~Yang,
  Eur. Phys. J. C {\bf 23}, 275 (2002).


\bibitem{Qian:2009dc}
  W.~Qian and B.~-Q.~Ma,
  Eur. Phys. J. C {\bf 65}, 457 (2010).


\bibitem{OTW97}
H.~B.~O'Connell, A.~W.~Thomas and A.~G.~Williams,
Nucl. Phys. {\bf A623}, 559 (1997).

\bibitem{ZLL13}
  Z.~-T.~Zou, C.~Li and C.~-D.~L¨¹,
  Chin.\ Phys.\ C {\bf 37}, 093101 (2013).





\bibitem{LHCbCPA}
  RAaij {\it et al.}  [LHCb Collaboration],
  JHEP {\bf 1306}, 112 (2013).

\bibitem{ddbar}
  G.~Isidori, J.~F.~Kamenik, Z.~Ligeti and G.~Perez,
  Phys. Lett. B {\bf711}, 46 (2012).



\bibitem{neutralCPA}
  S.~Blusk [LHCb Collaboration],
  arXiv:1212.4180 [hep-ex].




\end{thebibliography}
\end{document}